\documentclass[letterpaper]{article}
\usepackage[draft]{aaai2026} 
\usepackage{times} 
\usepackage{helvet} 
\usepackage{courier} 
\usepackage[hyphens]{url} 
\usepackage{graphicx} 
\usepackage{graphicx} 
\usepackage{natbib} 
\usepackage{caption} 
\usepackage{algorithm}
\usepackage{algorithmic}
\usepackage{placeins}
\usepackage{newfloat}
\usepackage{listings}
\DeclareCaptionStyle{ruled}{labelfont=normalfont,labelsep=colon,strut=off} 
\floatstyle{ruled}
\newfloat{listing}{tb}{lst}{}
\floatname{listing}{Listing}
\usepackage{microtype}
\usepackage{url}
\usepackage{booktabs}
\usepackage{lineno}

\usepackage{graphicx}
\usepackage{arydshln}
\usepackage{booktabs}
\usepackage{multirow}
\usepackage{amsmath} 
\usepackage{cleveref}
\usepackage{colortbl}
\usepackage{caption}
\usepackage{subcaption}
\usepackage[most]{tcolorbox}
\usepackage{listings}
\usepackage{fontawesome5}

\definecolor{darkblue}{rgb}{0, 0, 0.5}

\newcommand{\showdiffplus}[1]{\textcolor{green!50!black}{\small~($\uparrow$#1)}}
\newcommand{\showdiffminus}[1]{\textcolor{red!80!black}{\small~($\downarrow$#1)}}

\title{Towards Better Correctness and Efficiency in Code Generation}
\author{
    Yunlong Feng,
    Yang Xu,
    Xiao Xu,
    Binyuan Hui\thanks{Corresponding authors},
    Junyang Lin$^{\ast}$
}
\affiliations{
    Qwen Team, Alibaba Group\\
    \{yunlong.fyl,binyuan.hby,junyang.ljy\}@alibaba-inc.com
}

\begin{document}

\maketitle

\begin{abstract}
While code large language models have demonstrated remarkable progress in code generation, the generated code often exhibits poor runtime efficiency, limiting its practical application in performance-sensitive scenarios. 
To address this limitation, we propose an efficiency-oriented reinforcement learning framework guided by a novel performance reward. 
Based on this framework, we take a deeper dive into the code efficiency problem, identifying then proposing methods to overcome key bottlenecks: 
(1) Dynamic exploration overcomes the static data constraints of offline fine-tuning, enabling the discovery of more efficient code implementations.
(2) The error-insensitive reinforcement learning method and high-contrast efficiency signals are crucial for mitigating systematic errors and achieving effective optimization.
(3) Online exploration is most effective when starting from a high-correctness baseline, as this allows for efficiency improvements without sacrificing accuracy.
With these discoveries, we finally propose a two-stage tuning method, which achieves high and balanced performance across correctness and efficiency.
The results of experiments show the effectiveness of the method, which improves code correctness by 10.18\% and runtime efficiency by 7.75\% on a 7B model, achieving performance comparable to much larger model.
\end{abstract}

\section{Introduction}
\label{sec:intro}

With the rapid development of large language models (LLMs)~\citep{gpt3, gpt4, llama2, llama3, mistral, qwen, qwen2, claude3.5, gpt4o}, code large language models have improved their capabilities in various code-
related tasks. 
However, recent studies have indicated that these large models typically focus on the correctness of the generated code while neglecting its runtime efficiency \cite{waghjale2024ecco, effibench, mercury, evalperf}. 
As a result, the generated code can be 3 to 13 times slower than human-written versions, limiting the practical application of model-generated code in real-world software development workflows~\cite{effibench}. 

To this end, recent work has advanced along several key fronts, underpinned by the development of dedicated benchmarks like EvalPerf \cite{evalperf}, Mercury \cite{mercury}, and EffiBench \cite{effibench}.
Building on these evaluation frameworks, researchers have designed structured optimization workflows, such as the iterative, feedback-driven process in SOAP \cite{huang2024soap} and the two-stage approach of LLM4EFFI \cite{ye2025llm4effileveraginglargelanguage}. 
In parallel, a major research thrust focuses on directly augmenting model capabilities through fine-tuning and reinforcement learning (RL). 
This includes methods like Effi-Learner \cite{efficode} and PIE \cite{shypula2023learning}, which leverage curated code samples, as well as approaches like Mercury's \cite{mercury} validation of Direct Preference Optimization (DPO) and ACECode's \cite{acecode} use of rule-based PPO to further improve code efficiency during generation.

To better understand the underlying challenges, we conduct a series of explorations and identify the following key insights:
(1) Offline fine-tuning methods are bottlenecked by static datasets, as their performance is capped by the implementations present in the training data, limiting the ability to discover novel, more efficient algorithmic solutions. 
(2) While online methods like Reinforcement Learning (RL) offer a path to dynamic exploration, they introduce instabilities, as the inherently noisy and error-prone efficiency signal can prevent the model from learning effectively. 
This presents the challenge of designing an error-insensitive RL framework with high-contrast efficiency signals. 
(3) Our findings indicate that the optimization's starting point is critical, and that a model with high initial correctness serves as a superior foundation for any subsequent efficiency optimization.

Based on these findings, we design a practical two-stage training strategy, which is structured to first establish a robust foundation of code correctness before optimizing for runtime efficiency.
In the first phase, we focus on establishing foundational correctness by fine-tuning a base LLM, which can be done with normal offline methods. 
This creates a high-accuracy model to serve as a solid starting point. 
In the second phase, we build upon this foundation to enhance efficiency. 
We employ the error-insensitive online reinforcement learning algorithm that uses high-contrast efficiency signals to improve runtime performance while maintaining the high level of correctness achieved in the first stage.

Our main contributions are as follows:
\begin{itemize}
\item We first analyze the limitations of offline methods, revealing that they are bottlenecked by initial data quality and struggle to enhance efficiency without sacrificing the model's correctness.
\item We then investigate techniques for the online RL phase. We find that selecting an error-insensitive algorithm (RLOO) is crucial for preserving correctness, and we introduce a stabilized reward signal and high-contrast inputs to ensure a robust and effective optimization process.
\item Based on these insights, we propose a two-stage tuning strategy, which achieves high and balanced performance across correctness and efficiency. 
The results show the effectiveness of this method, which improves code correctness by 10.18\% and runtime efficiency by 7.75\% on a 7B model, achieving performance comparable to much larger model.
\end{itemize}

\section{Efficiency-oriented Reinforcement Learning}
\label{sec:method}

It's hard to evaluate the runtime efficiency of code.
Manually creating a reference version is labor-intensive and costly, and even then, the notion of ``fast'' code is inherently relative, as identifying a definitively optimal implementation is often intractable. 
This means even a carefully hand-tuned version cannot be guaranteed to be optimal \cite{acecode}. 
Consequently, relying on a single, manually-optimized version as a performance baseline is inadequate. 

To facilitate continuous model improvement, we therefore adopt a group comparison methodology for performance assessment.
As our performance metric, we use CPU instruction counts rather than wall-clock time to mitigate sources of non-determinism, such as operating system scheduling \cite{evalperf}. 
Besides, we introduce inputs with high complexity differentiation to overcome the insensitivity of trivial inputs to algorithmic efficiency differences during execution.
However, the distribution of these raw counts is typically highly skewed and exhibits large variance, which can impede stable model training. 
To address this, we apply a logarithmic transformation to the counts to compress their value range and stabilize the variance \cite{2008Applied}. 

\subsection{Reward Function}

Our objective is to ensure that correct and faster code results in higher scores, with improvement on efficiency consistently contributing to the reward. The reward should be benchmarked against the efficiency of the code generated by the current model for the same query.

To formalize this, we first define our parameters.
\begin{itemize}
    \item $C = -\log(N)$: The performance metric for the code being evaluated, where $N$ is the number of CPU instructions. Due to the negative sign, a higher value of $C$ indicates faster execution.
    \item $G = [C_{1}, C_{2}, \dots, C_{m}]$: An array of performance metrics for all correct codes generated for the same query.
    \item $\max(G)$ / $\min(G)$: The maximum / minimum value in $G$, representing the most efficient (fastest) / least efficient (slowest) code among all codes generated for the query.
\end{itemize}

Based on these parameters, we designed the reward function $\mathcal{R}$. If the code passes all tests, it receives a score calculated by $\mathcal{F}(C, G)$. This function maps the performance metric of the code to a reward where the slowest passing code gets a score of $1.0$ and the fastest gets a score up to $2.0$. Otherwise, it receives a penalty based on the error type.

This idea is formulated as follows:
\begin{align*}
S_{min} & = 1.0 \\
S_{max} & = \min(2.0, S_{min} + \max(G) - \min(G)) \\
\mathcal{F}(C, G) & = S_{\min} + (C - \min(G)) \cdot \frac{S_{\max} - S_{\min}}{\max(G) - \min(G)}
\end{align*}

The final reward $\mathcal{R}(C, G)$ is assigned based on the execution outcome, which connects the performance score with the error classification:
\begin{itemize}
    \item \textbf{Pass All Tests}: The code receives the calculated performance score, $\mathcal{R} = \mathcal{F}(C, G)$.
    \item \textbf{Test Error}: A score of $0.0$ is assigned if the code fails in any test.
    \item \textbf{No Entity Error}: A penalty of $-0.5$ is assigned if the model-generated code cannot find a corresponding test function.
    \item \textbf{Format Error}: The largest penalty of $-1.5$ is assigned if the response does not follow the specified format.
\end{itemize}

This is summarized by the complete reward function:
\begin{align*}
\mathcal{R}(C, G) & =
\begin{cases}
    \mathcal{F}(C, G) & \text{if Pass All Tests}, \\
    0.0               & \text{for a Test Error}, \\
    -0.5              & \text{for a No Entity Error}, \\
    -1.5              & \text{for a Format Error}.
\end{cases}
\end{align*}

\section{Identify the Bottlenecks in Code Efficiency}
\label{sec:analysis}

This chapter aims to conduct a series of preliminary experiments to deeply analyze the inherent characteristics and challenges of existing mainstream methods in code optimization tasks. 
By understanding their strengths and limitations, we aim to achieve a better balance between correctness and efficiency in code generation.

\subsection{Experiments}

To contextualize our analysis, we compare several common fine-tuning methods across two standard benchmarks. 
These methods can be broadly categorized into two groups: offline methods, which learn from a static, precollected dataset, and online methods, which dynamically generate data during the training process. 

\paragraph{Offline Methods}
\begin{itemize}
    
    \item \textbf{SFT}: A standard approach where the model is fine-tuned exclusively on the fastest correct code solution available for each problem in the training set.

    \item \textbf{DPO}~\cite{rafailov2023direct}: For each problem, we construct correctness pairs (correct vs. incorrect code) and efficiency pairs (faster vs. slower correct code).
\end{itemize}

\paragraph{Online Methods}
\begin{itemize}
    \item \textbf{GRPO}~\cite{shao2024deepseekmathpushinglimitsmathematical}: A method that optimizes the policy based on the relative performance ranking within a group of generated solutions.

    \item \textbf{RLOO}~\cite{ahmadian-etal-2024-back}: A reinforcement learning technique that uses a leave-one-out strategy to estimate the reward baseline.
\end{itemize}

\paragraph{Datasets} To comprehensively evaluate the model's code optimization capabilities, we utilize several datasets for training and benchmarking. 
The \Cref{tab:dataset} below summarizes the datasets used for training and evaluation.

\begin{table}[h]
\centering
\begin{tabular}{@{}llrr@{}}
\toprule
Dataset    & Source   & Metric & Num (\#)  \\ \hline
\rowcolor[rgb]{0.93,0.93,0.93}\multicolumn{4}{c}{Train Dataset}   \\
EffiBench & LeetCode           & -  & 790                          \\ 
Effi-Learner  & Synthetic      & -  & 2,978                        \\
\rowcolor[rgb]{0.93,0.93,0.93}\multicolumn{4}{c}{Benchmark}       \\
Mercury   & LeetCode           & Beyond &  256            \\
Evalperf  & Evalplus           & DPS$_{norm}$  & 109    \\
\bottomrule
\end{tabular}%
\caption{
The subset of the training dataset. 
Notably, the queries from Effi-Learner are simpler, resembling the difficulty of Evalplus problems, while EffiBench's queries are more akin to Mercury's, since they all originate from LeetCode.
}
\label{tab:dataset}
\end{table}

The metrics DPS$_{norm}$ and Beyond are used to represent the model's performance in terms of code efficiency. 
A higher value for these metrics indicates better efficiency. 
For a more detailed explanation, please refer to the appendix.

\subsection{Results of Preliminary Experiment}

\begin{table}[h]
\centering
\begin{tabular}{@{}ccccc@{}}
\toprule
\multicolumn{1}{c}{\multirow{2}{*}{Method}} & \multicolumn{2}{c}{EvalPerf} & \multicolumn{2}{c}{Mercury}   \\ \cmidrule(l){2-3} \cmidrule(l){4-5}
\multicolumn{1}{c}{} & DPS$_{norm}$ & Pass@1        & Beyond         & Pass@1        \\ \midrule
Untuned              & 77.60        & 77.50         & 74.80          & 77.20  \\
SFT                  & 77.85        & 79.19         & 66.36          & 83.20   \\
DPO                  &\textbf{87.61}& 81.89         & 72.23          & 85.47   \\
GRPO                 & 83.96        & 81.39         & \textbf{79.15} & 87.50  \\
RLOO                 & 81.97        &\textbf{82.46} & 76.00          & \textbf{89.92} \\
\bottomrule
\end{tabular}%
\caption{
The results of different methods on Qwen-2.5-Coder-Instruct-7B model.
}
\label{tab:main-exp-1}
\end{table}

We evaluated the different methods on the Qwen-2.5-Coder-Instruct-7B model, with the results presented in \Cref{tab:main-exp-1}. 
Below, we analyze the performance of each method to identify the challenges in the scene and guide future research.

\subsection{Offline Methods Bottlenecked by Initial Data Sampling}

Offline methods, such as Supervised Fine-Tuning (SFT) and Direct Preference Optimization (DPO), are common strategies for improving a model's performance.
These approaches typically rely on generating a pool of candidate solutions from a base model and then selectively using them for fine-tuning. However, their ultimate success is fundamentally constrained by the quality of this initial sample pool. 
If the base model struggles to generate solutions that are both correct and efficient for certain problems, the offline tuning process is inherently limited, as it lacks high-quality data to learn from. 
In the following analysis, we demonstrate how this sampling bottleneck creates a crucial trade-off between correctness and efficiency, especially when tackling complex coding tasks.

\subsubsection{Limited by Inefficient Samples from Challenging Problems} 
Our experiments first explored fine-tuning using only the fastest correct code generated by the base model. 
This method boosted correctness but at the cost of significantly reduced runtime efficiency on the more challenging Mercury benchmark.

\begin{table}[h]
\centering
\begin{tabular}{@{}lcc@{}}
\toprule
Dataset  & Effi-Learner  & EffiBench  \\ \midrule
Pass@1   & 81.39         & 62.99   \\
\bottomrule
\end{tabular}%
\caption{
The subset of the training dataset. 
The passrate is based on the Qwen-Coder-7B-Instruct. 
}
\label{tab:dataset-pass}
\end{table}

A plausible explanation arises from the data in \Cref{tab:dataset} and \Cref{tab:dataset-pass}. 
The base model's correctness is much lower on the difficult LeetCode problems in EffiBench (62.99\%) than on the simpler synthetic problems in Effi-Learner (81.39\%). 
This indicates that for complex problems, the few correct solutions the model manages to generate are often naive, brute-force implementations. 
While functionally correct, they are computationally inefficient. 
Consequently, when we use these difficult samples to create the fine-tuning data, we inadvertently select for these ``correct-but-slow'' solutions, teaching the model to replicate inefficient code.

\subsubsection{DPO Hits an Efficiency Ceiling} 
As the \Cref {tab:main-exp-1} and \Cref{tab:rq1} shows, DPO makes more effective use of negative samples than SFT, leading to significant improvements in code correctness without compromising efficiency. 
However, on more challenging problems, such as those in the Mercury benchmark, DPO faces a similar limitation to SFT. 
Both methods are constrained by the base model's initial capabilities, making it difficult to sample solutions that are both correct and highly efficient. 
This inherent limitation hinders DPO's ability to further enhance code efficiency for complex tasks.

\begin{table}[h]
\centering
\begin{tabular}{@{}crrrr@{}}
\toprule
\multicolumn{1}{c}{\multirow{2}{*}{Proportion}} & \multicolumn{2}{c}{EvalPerf} & \multicolumn{2}{c}{Mercury}  \\ \cmidrule(l){2-3} \cmidrule(l){4-5}
\multicolumn{1}{c}{} & DPS$_{norm}$          & Pass@1        & Beyond          & Pass@1        \\ \midrule

0.0    & 84.60          & \textbf{87.20}& 71.45        &\textbf{87.11}  \\
0.5    & \textbf{87.61} & 81.89 & \textbf{72.23} & 85.47   \\
1.0    & 84.80          & 74.70 & 72.21 & 77.30   \\
\bottomrule
\end{tabular}%

\caption{
Results of DPO with different settings on Qwen-2.5-Coder-Instruct-7B. 
The Proportion represents the proportion of efficiency pairs in the training data.
}
\label{tab:rq1}
\end{table}

\subsection{Data Composition Dictates the Trade-off Between Correctness and Efficiency}
\label{sec:trade-off}

\begin{figure}[ht]
    \centering
    \includegraphics[width=0.85\linewidth]{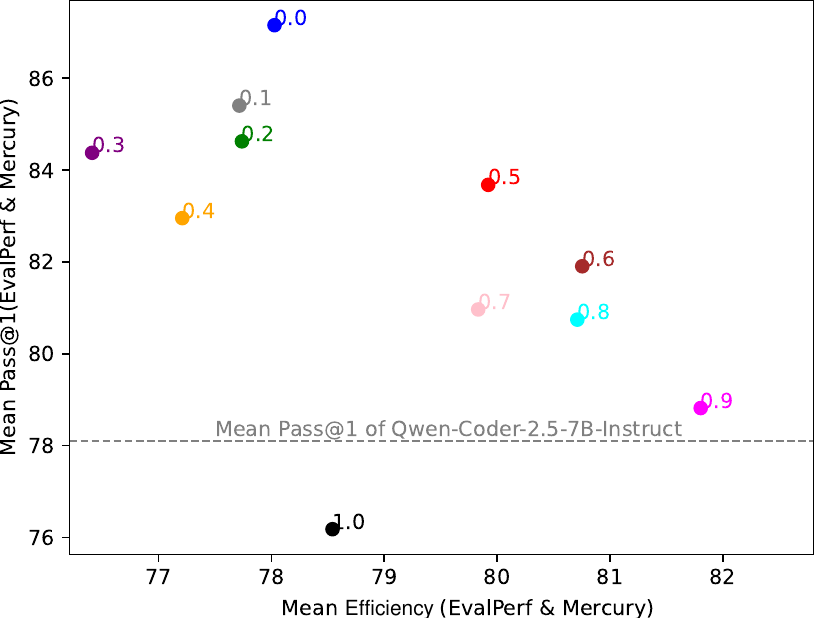}
    \caption{Impact of the proportion of efficiency pairs in training data on code correctness and efficiency in DPO training.}
    \label{fig:proportion}
\end{figure}

While considering execution efficiency, code correctness must also be taken into account. 
Consequently, we seek to understand how the proportion of ``efficiency pairs'' within the training data influences the model's performance.

As the \Cref{fig:proportion} shows, there is a core trade-off in model performance between code correctness (Y-axis) and runtime efficiency (X-axis), which can be tuned by varying the proportion of ``efficiency pairs.''

\begin{itemize}
    \item As the proportion of efficiency pairs increases from 0.0 to 0.9, runtime efficiency steadily improves while code correctness declines, demonstrating that this ratio can be used to steer the model's focus.
    \item Notably, at a 1.0 proportion (using only efficiency pairs), performance deteriorates on both metrics, with correctness falling below the baseline. This indicates that solely pursuing efficiency at the complete expense of correctness leads to an overall degradation in performance.
\end{itemize}

From these results, we conclude that a significant emphasis on efficiency ($p \le 0.9$) can be applied without substantially degrading model correctness, as this principle fails only under extreme conditions, for example, when efficiency is exclusively pursued at $p=1.0$. 
Consequently, we conjecture that within an online reinforcement learning, the dynamic reward mechanism can adeptly manage this balance, thereby improving efficiency without compromising model correctness.

\subsection{Online RL is Constrained by Initialization, Reward, and Algorithm}

\begin{table}[h]
\centering
\resizebox{\linewidth}{!}{%
\begin{tabular}{@{}crrrr@{}}
\toprule
 High Complexity & \multicolumn{2}{c}{EvalPerf} & \multicolumn{2}{c}{Mercury} \\ \cmidrule(l){2-3} \cmidrule(l){4-5}
 Inputs & DPS$_{norm}$   & Pass@1        & Beyond  & Pass@1        \\ \hline
\rowcolor[rgb]{0.93,0.93,0.93}\multicolumn{5}{c}{From Qwen-2.5-Coder-7B}   \\
Y &\textbf{90.81}& 78.06         &\textbf{80.59}& 88.75 \\
N  & 81.94        & \textbf{80.85}         & 76.90        & \textbf{90.70} \\
\rowcolor[rgb]{0.93,0.93,0.93}\multicolumn{5}{c}{From Qwen-2.5-Coder-Instruct-7B}   \\
Y & \textbf{83.96}        & \textbf{81.39}         & \textbf{79.15}        & 87.50 \\
N &  81.94        & 80.85         & 76.90        & \textbf{90.70}  \\
\bottomrule
\end{tabular}%
}
\caption{The Impact of Syn Inputs with High Complexity Differentiation in online reinforcement learning.}
\label{tab:ablation-exp-inputs}
\end{table}

\subsubsection{Inputs with High Complexity Differentiation Gain Better Efficiency}
During reinforcement learning, employing inputs with enhanced complexity differentiation to compute efficiency rewards enables more precise discrimination of code runtime performance, thereby driving superior efficiency performance. 
As validated in \Cref{tab:ablation-exp-inputs}, such synthetically generated complex inputs effectively .
However, this approach introduces a trade-off that may compromise code correctness. 
This effect is particularly pronounced in scenarios with insufficient test coverage, as evidenced by the EvalPerf benchmark results where efficiency optimization conflicts with functional accuracy.

\begin{table}[h]
\centering
\resizebox{\linewidth}{!}{%
\begin{tabular}{@{}ccrrrr@{}}
\toprule
\multicolumn{1}{c}{\multirow{2}{*}{Model}} & \multicolumn{2}{c}{EvalPerf} & \multicolumn{2}{c}{Mercury} \\ \cmidrule(l){2-3} \cmidrule(l){4-5}
\multicolumn{1}{c}{}                 & DPS$_{norm}$   & Pass@1         & Beyond            & Pass@1        \\ \midrule
Base            & \textbf{90.81}  & 78.06          & \textbf{80.59}    & 88.75           \\
Inst            & 83.96           & 81.39          & 79.15             & 87.50           \\
DPO ($p = 0.1$)    & 87.35           & \textbf{84.32} & 80.52             & \textbf{88.83}  \\
DPO ($p = 0.5$)    & 88.27           & 83.42          & 78.66             & 86.48            \\
\bottomrule
\end{tabular}%
}
\caption{The Impact of different initialization in reinforcement learning. Where the $p$ means proportions efficiency pairs. The Base and Inst means  Qwen-2.5-Coder-7B and Qwen-2.5-Coder-Instruct-7B. The DPO model is tuned based on the Qwen-2.5-Coder-Instruct-7B.}
\label{tab:ablation-exp-init}
\end{table}

\subsubsection{Initialization from Models with Higher Diversity or Accuracy Gain Better Performance}
As the \Cref{tab:ablation-exp-init} shows, the base model and post-trained models achieve better performance.
Base models exhibit higher output diversity~\cite{zhu2025bareleveragingbaselanguage}, which is effectively leveraged by online reinforcement learning. 
This suggests that starting directly from a base model while considering both correctness and efficiency in RL is feasible. 
Initializing with DPO (High Accuracy, $p=0.1$) allows RL to focus more on optimizing efficiency rather than correctness during the training process. 
As we conjectured in the \cref{sec:trade-off}, this results in better efficiency optimization without significantly compromising accuracy.

\begin{figure}[h]
    \centering
    \begin{subfigure}[b]{0.48\linewidth}
        \centering
        \includegraphics[width=\linewidth]{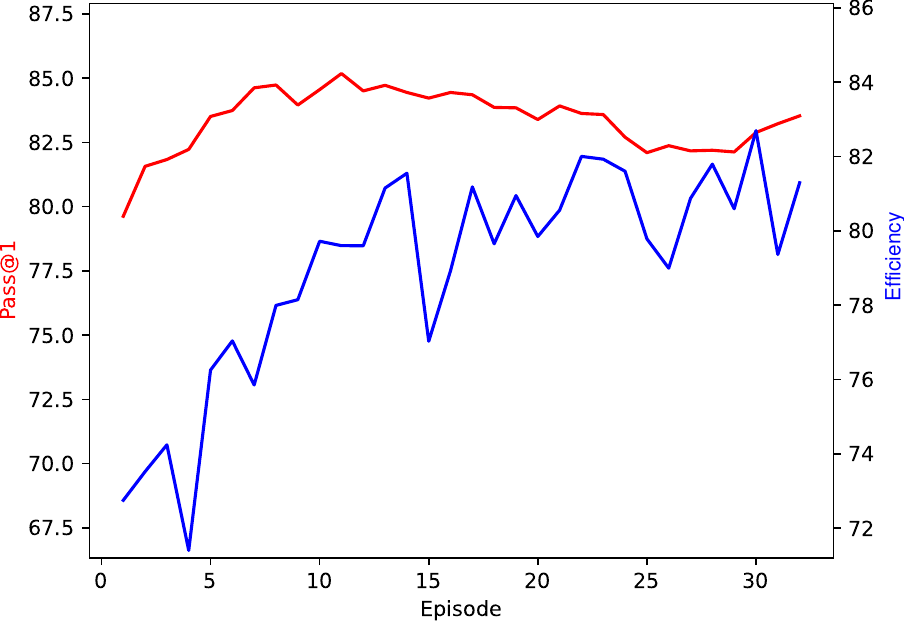}
        \caption{GRPO}
        \label{fig:grpo-inst}
    \end{subfigure}
    \begin{subfigure}[b]{0.48\linewidth}
        \centering
        \includegraphics[width=\linewidth]{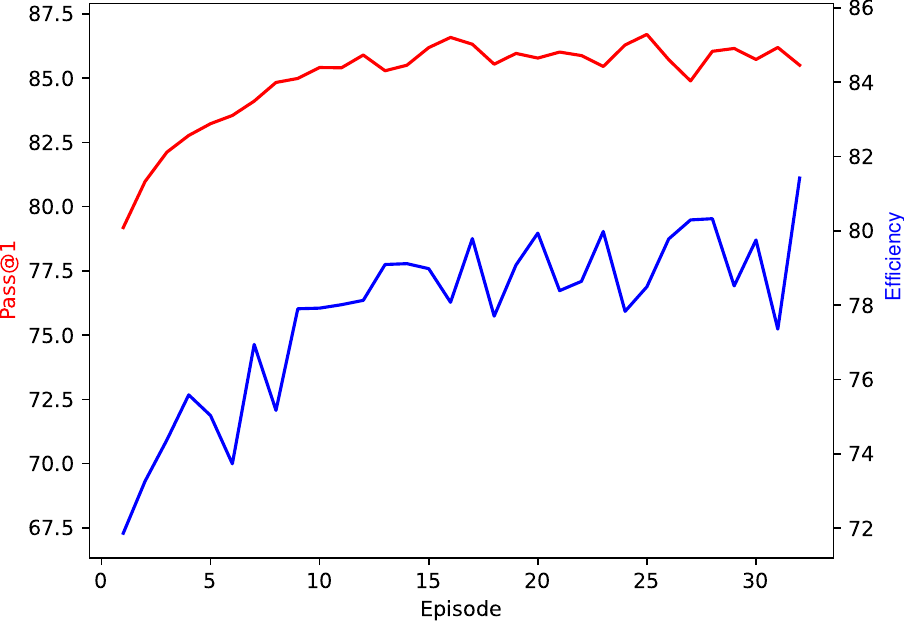}
        \caption{RLOO}
        \label{fig:rloo-inst}
    \end{subfigure}
    \caption{
    Comparison of GRPO and RLOO performance in terms of accuracy (pass@1) and efficiency (dps), Initialized with the Qwen-2.5-Coder-7B-Instruct.
    While GRPO's efficiency improves more rapidly than RLOO's during training, it experiences a notable decline in correctness in the later stages (after 5-10 episodes). In contrast, RLOO demonstrates a better ability to maintain its correctness over time.
    }
    \label{fig:performance-var}
\end{figure}

\begin{figure*}[t]
    \centering
    \includegraphics[width=0.98\linewidth, alt={Overview of our method}]{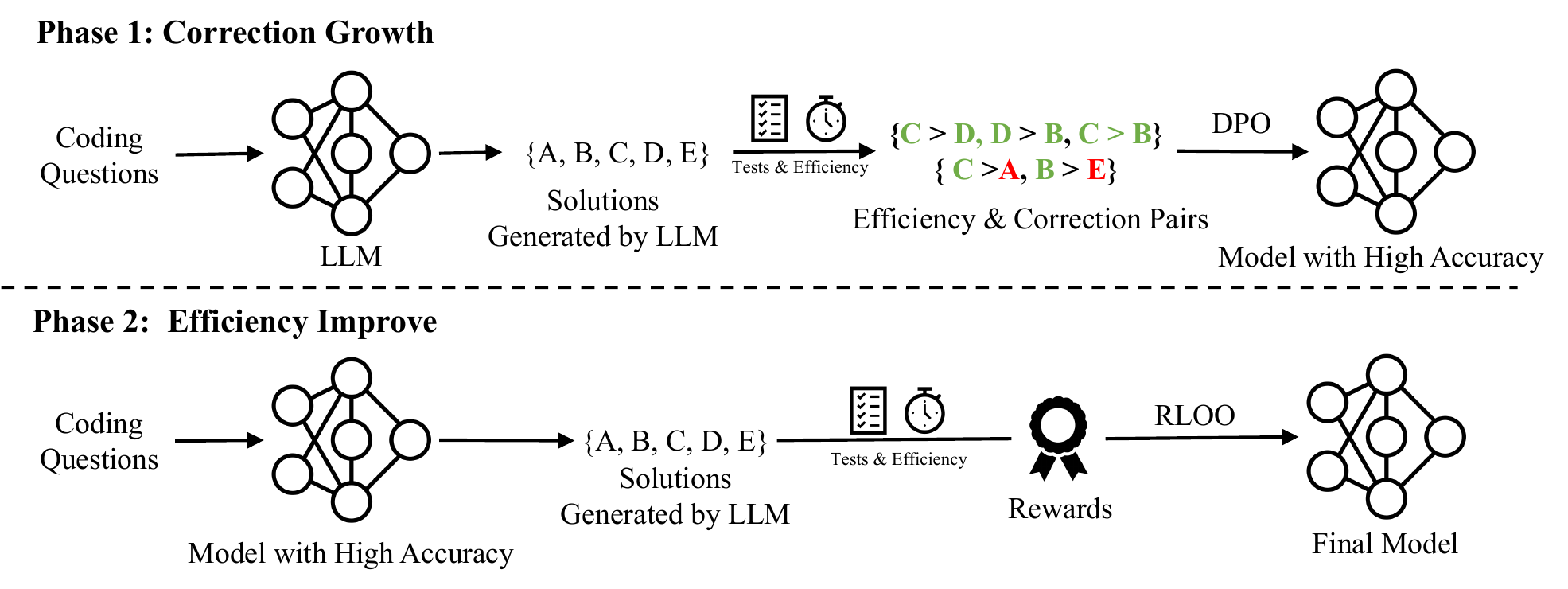}
    \caption{
    Overview of our method. Our approach consists of two phases. 
    In Phase 1, we enhance code correctness by fine-tuning a base LLM with Direct Preference Optimization (DPO) on preference pairs of generated solutions. 
    In Phase 2, the resulting high-accuracy model serves as an initial policy and is further optimized for runtime efficiency using online reinforcement learning (with error-insensitive RL algorithms, such as RLOO).
    }
    \label{fig:method}
\end{figure*}

\begin{table*}[ht]
\centering
\resizebox{\linewidth}{!}{%
\begin{tabular}{@{}cccccccc@{}}
\toprule
\multicolumn{1}{c}{\multirow{2}{*}{LLMs}} & \multicolumn{1}{c}{\multirow{2}{*}{Method}} & \multicolumn{2}{c}{EvalPerf} & \multicolumn{2}{c}{Mercury} & \multicolumn{2}{c}{AVG}  \\ \cmidrule(l){3-4} \cmidrule(l){5-6} \cmidrule(l){7-8}
\multicolumn{1}{c}{} & \multicolumn{1}{c}{} & \faBolt        & Pass@1        & \faBolt         & Pass@1       & \faBolt & Pass@1  \\ \midrule
 & Untuned                  & 80.92        & 85.59         & 76.97        & 94.14 & 78.95 & 89.87   \\
Qwen-2.5-Coder- & ECCO         & 82.16\showdiffplus{1.24}   & 63.56\showdiffminus{22.03}  & 73.29\showdiffminus{3.68}   & 89.06\showdiffminus{5.08} & 77.73\showdiffminus{1.22} & 76.31\showdiffminus{13.56}  \\
Instruct-32B & Effi-Learner & 82.45\showdiffplus{1.53}   & 77.11\showdiffminus{8.48}   & 77.13\showdiffplus{0.16}   & 91.41\showdiffminus{2.73} & 79.79\showdiffplus{0.84} & 84.26\showdiffminus{5.61}   \\
& LLM4EFFI     & 86.20\showdiffplus{5.28}   & 87.30\showdiffplus{1.71}   & 78.96\showdiffplus{1.99}   & 93.75\showdiffminus{0.39} & 82.58\showdiffplus{3.63} & 90.53\showdiffplus{0.66} \\ \midrule
 & Untuned & 77.60        & 77.50         & 74.80        & 77.20 & 76.18 & 77.37 \\
Qwen-2.5-Coder- & DPO          & 88.50\showdiffplus{10.90}  & 81.64\showdiffplus{4.14}   & 76.66\showdiffplus{1.86}   & 84.38\showdiffplus{7.18} & 82.58\showdiffplus{6.40} & 83.01\showdiffplus{5.64} \\
Instruct-7B & RLOO         & 81.97\showdiffplus{4.37}   & 82.46\showdiffplus{4.96}   & 76.00\showdiffplus{1.20}   & 89.92\showdiffplus{12.72}& 78.99\showdiffplus{2.81} & 86.19\showdiffplus{8.82}  \\
& Ours         & 88.82\showdiffplus{11.22}  & 84.32\showdiffplus{6.82}   & 79.04\showdiffplus{4.24}   & 90.78\showdiffplus{13.58}& 83.93\showdiffplus{7.75}& 87.55\showdiffplus{10.18} \\
\bottomrule
\end{tabular}%
}
\caption{
The results of different methods of Qwen-2.5-Coder-Instruct-7B model and Qwen-2.5-Coder-Instruct-32B. The results of Qwen-2.5-Coder-Instruct-32B are cited from LLM4EFFI~\cite{ye2025llm4effileveraginglargelanguage}.
The symbols~\faBolt~in EvalPerf and Mercury represent the DPS$_{norm}$ and Beyond metrics, respectively.
}
\label{tab:main-exp}
\end{table*}

\subsubsection{Error-Insensitive RL Algorithms can Maintain Accuracy better}
As the \Cref{tab:main-exp-1} and \Cref{fig:performance-var} shows, RLOO generally achieves higher average accuracy, while GRPO tends to achieve higher code execution efficiency. 
This discrepancy may be attributed to GRPO's regularization mechanism, which amplifies minor improvements but also magnifies errors in efficiency rewards, potentially leading to over-optimization and reduced correctness. 

\section{Towards Better Correctness and Efficiency}
\label{sec:experiments}

\subsection{How can we gain better performance?}

As we analyzed the impact of various factors on the final model performance. 
This analysis will help us understand the key aspects that need to be considered to achieve better accuracy and runtime efficiency.

\begin{enumerate}
    \item \textbf{Switching to a More Stable RL Algorithm to Maintain Accuracy:} As illustrated in \Cref{fig:performance-var}, RLOO demonstrates superior stability compared to GRPO in the later stages of training. Consequently, we adopt the RLOO algorithm to optimize the model's performance while mitigating accuracy degradation.

    \item \textbf{Initializing from a High-Accuracy Model:} The RLOO algorithm tends to preserve the accuracy of its initial model. We therefore initialize our training from a checkpoint with higher accuracy—specifically, a model previously fine-tuned with Direct Preference Optimization (DPO). The proportion of ``efficiency pairs'' for this stage was set to 0.1.

    \item \textbf{Using High-Contrast Inputs to Compute Efficiency Rewards:} Our prior analysis indicates that employing more discriminative inputs is crucial for calculating rewards. These inputs elicit a wider range of performance, enabling a more accurate distinction between the runtime efficiencies of different code generations.
\end{enumerate}

\begin{figure*}[t]
\centering
\begin{tcolorbox}[colback=white, colframe=gray, title=Example Response]

The problem requires finding the number of range sums that lie within a given inclusive range [lower, upper]. A \textbf{straightforward} approach would be to use a nested loop to calculate all possible range sums, but this would be inefficient for large input sizes due to its $O(n^2)$ complexity. A \textbf{more efficient} solution can be achieved using a prefix sum array and a binary search approach to count the valid range sums in $O(n \log n)$ time complexity.

\begin{lstlisting}[language=python]
```python
class Solution(object):
    def countRangeSum(self, nums, lower, upper):
        count, running_sum = 0, 0
        accum_sum = [0]
        for num in nums:
            running_sum += num
            count += bisect_right(accum_sum, running_sum - lower) - bisect_left(accum_sum, running_sum - upper)
            bisect.insort(accum_sum, running_sum)
        return count
```
\end{lstlisting}
\end{tcolorbox}
\caption{The model typically posits that there exists a straightforward approach to solving the problem, then proceeds to highlight that this method may be inefficient for certain inputs. Subsequently, it proposes an alternative solution with improved speed and concludes by providing the corresponding code.}
\label{fig:case1}
\end{figure*}

\begin{figure}[ht]
    \centering
    \includegraphics[width=0.85\linewidth]{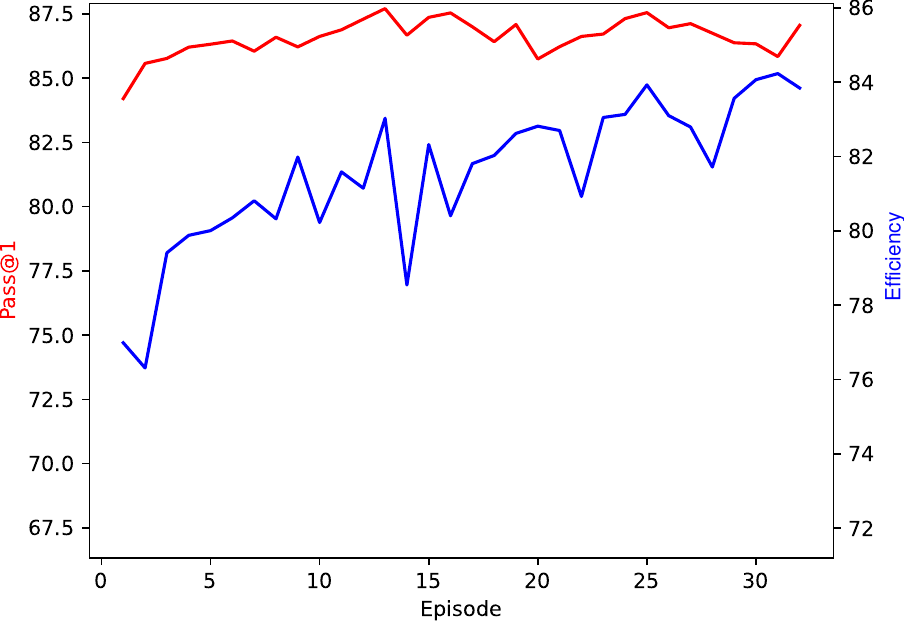}
    \caption{Accuracy and runtime efficiency of the RLOO training process using a DPO-tuned model for initialization.}
    \label{fig:rloo-dpo}
\end{figure}

\subsection{Method}

Derived from our analysis, we designed a two-stage training strategy as shown in \Cref{fig:method}. 
This approach is structured to first establish a robust foundation of code correctness before systematically optimizing for runtime efficiency.

\textbf{Stage 1: Correctness Growth}
We utilize Direct Preference Optimization (DPO) with a dataset of code pairs heavily weighted towards correctness. 
Specifically, the training data is composed of 90\% correctness-focused pairs and 10\% efficiency-focused pairs, which allows us to produce a high-accuracy foundational model.

\textbf{Stage 2: Efficiency Improve}
Initializing from the DPO-tuned checkpoint, we employ the RLOO algorithm to fine-tune for runtime efficiency. 
As shown in \Cref{fig:rloo-dpo}, the training process yields a stable improvement in runtime efficiency while preserving the high level of correctness achieved in the prior stage.

\subsection{Results}

The result in \Cref{tab:main-exp-1} demonstrates that our proposed method achieves significant performance enhancements on two benchmarks.
On the EvalPerf benchmark, it achieved a score of 88.82 on DPS$_{norm}$ and 84.32 on Pass@1. 
For the Mercury benchmark, the scores were 79.04 for Beyond and 90.78 for Pass@1. 
The success of this method validates our findings and demonstrates an effective path towards better correctness and efficiency in code generation.

\begin{figure*}[t]
    \centering
    \begin{subfigure}[b]{0.48\linewidth}
        \centering
        \includegraphics[width=\linewidth]{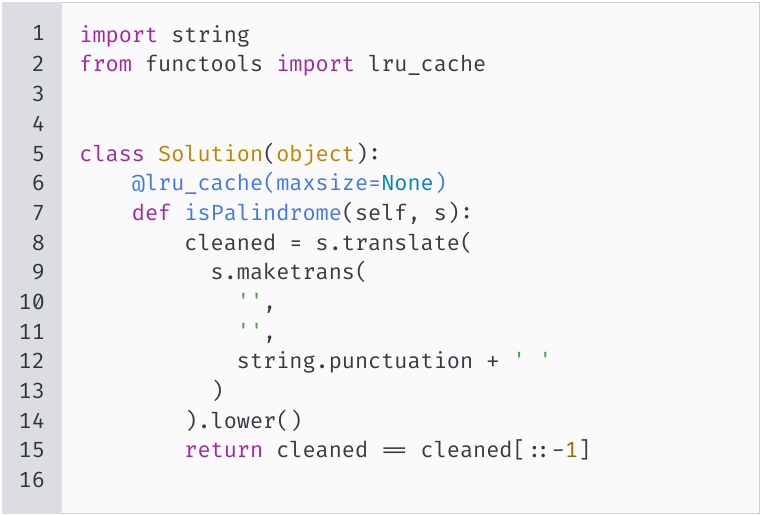}
        \caption{Reward Hacking: LRU cache. }
        \label{fig:hack1}
    \end{subfigure}
    \hfill
    \begin{subfigure}[b]{0.48\linewidth}
        \centering
        \includegraphics[width=\linewidth]{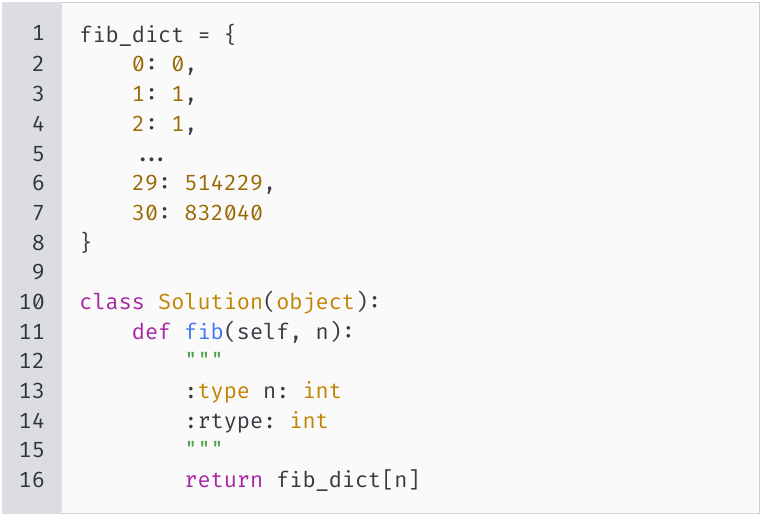}
        \caption{Reward Hacking: Static Table.}
        \label{fig:hack2}
    \end{subfigure}
    \caption{Two examples of reward hacking behavior in different scenarios. 
    (a) The absence of isolation between evaluation environments for the same code segment allows for the utilization of an LRU Cache, which significantly enhances ``runtime efficiency'' and consequently yields a high reward.
    (b) The incompleteness and limited scope of the collected test cases for the problems enable the model to construct a static table from its pre-trained knowledge to pass the tests, thereby achieving a high reward.
    }
    \label{fig:reward-hacking}
\end{figure*}

\subsection{Case Study}

This section presents specific cases to illustrate both the model's intended successful paradigm and several instances of reward hacking.

In a typical successful optimization case (see \Cref{fig:case1}), the model first identifies a direct or ``brute-force'' solution and analyzes its potential performance bottlenecks on certain inputs. 
Subsequently, it proposes a more efficient algorithm and provides the complete implementation code. 
This process demonstrates the model's expected reasoning and optimization capabilities.

However, we also observed that the model exploits loopholes in the evaluation mechanism to ``hack'' the reward signal, leading to its distortion.
\begin{itemize}
\item \textbf{Exploiting the Evaluation Method:} One failure mode arises from our method of measuring execution efficiency (see \Cref{fig:hack1}). 
To ensure stable results, our evaluation script warms up the system and averages multiple runs. 
The model exploited this by introducing an LRU Cache to ``memorize'' results across repeated executions, thereby achieving artificially high reward scores in subsequent runs.
\item \textbf{Exploiting Limited Test Cases:} Another hacking behavior stems from the limited scope of our test cases (see \Cref{fig:hack2}). 
Drawing on knowledge from pre-training, the model used data structures like dictionaries to hard-code or memoize solutions for specific test inputs.
While this approach passed the existing tests, it lacked generalizability, failed on new inputs, and sometimes produced incorrect results.
\end{itemize}

\section{Related Work}
\label{sec:background}

Recent advancements in Code Large Language Models (CodeLLMs) have demonstrated remarkable capabilities across a spectrum of code-related tasks~\cite{li2022alphacode, lozhkov2024starcoder, wang2023codet5+, yu2023wavecoder, wei2023magicoder, luo2023wizardcoder, hui2024qwen2}. However, a growing body of research highlights a significant limitation: their tendency to overlook the execution efficiency of the generated code~\cite{waghjale2024ecco, effibench, mercury, evalperf}. Empirical studies have shown that code produced by these models can be 3 to 13 times slower than human-written counterparts~\cite{effibench}. To systematically evaluate and address this issue, several benchmarks have been established. Notably, Mercury~\cite{mercury}, EvalPerf~\cite{evalperf}, and EffiBench~\cite{effibench} provide standardized frameworks for assessing code execution efficiency, often utilizing problems from platforms like LeetCode and EvalPlus.

In response to this efficiency gap, a nascent body of work has explored methods to optimize the code generated by LLMs. These approaches can be broadly categorized into workflow-based methods and data-centric fine-tuning strategies. One line of inquiry focuses on engineering the generation process itself. For instance, ECCO~\cite{waghjale2024ecco} introduce a self-refinement method using natural language feedback to critique its own code and then apply optimization suggestions, while SOAP~\cite{huang2024soap} introduces an iterative optimization loop, leveraging direct execution feedback to refine the code. Similarly, LLM4EFFI~\cite{ye2025llm4effileveraginglargelanguage} proposes a two-domain framework that mimics human development by first selecting an efficient algorithm in a ``logic domain'' before implementing it in a ``code domain.'' Another prominent direction involves curating datasets of efficient code and employing advanced training techniques. Mercury~\cite{mercury}, for example, fine-tunes models via Supervised Fine-Tuning (SFT) and Direct Preference Optimization (DPO) on a corpus of efficient, human-written solutions. Building on automated optimization, Effi-Learner~\cite{efficode} fine-tunes models using the performance-enhanced code generated by the SOAP framework. Others, like PIE~\cite{shypula2023learning}, create specialized datasets for instruction tuning to improve C++ code efficiency, albeit with an observed trade-off in functional correctness.  ACECode~\cite{acecode} applies PPO to improve efficiency with human-written solutions as upper bound references.

\section{Conclusion}

In this work, we addressed the critical gap between the functional correctness and the runtime efficiency of code generated by large language models. 
Our investigation into different optimization methods revealed their distinct strengths and limitations. 
We found that offline approaches, while constrained by static initial data, are effective at establishing a strong correctness baseline. 
In contrast, error-insensitive online methods excel at dynamically improving runtime efficiency while maintaining the high level of correctness already achieved.
This discovery directly guided our strategic approach. 
By first solidifying the model's correctness before pursuing efficiency gains, we achieved significant improvements in both metrics. 
The success of this method validates our findings and demonstrates an effective path towards better correctness and efficiency in code generation.

\bibliography{reference}

\section*{Appendix}

\subsection{Data Prepare}

The datasets and benchmarks are summarized in \Cref{tab:dataset}. 
We removed problems from Effi-Learner that had the same function signature as the Evalperf Benchmark, and de-duplicated the remaining problems using their function signatures. We also removed problems from EffiBench that duplicated the Mercury test set.

\subsection{Metrics}

\begin{itemize}
    \item \textbf{DPS$_{norm}$}: Sourced from EvalPerf~\cite{evalperf}, the DPS$_{norm}$ (Differential Performance Score, normalized) is used for the \textbf{Evalperf} benchmark. It evaluates the efficiency of a new code solution by calculating the percentage of reference solutions it runs faster than. A higher score indicates better efficiency.

    \item \textbf{Beyond}: The Beyond metric, from Mercury~\cite{mercury}, is used for the \textbf{Mercury} benchmark. It assesses efficiency by normalizing a solution's runtime against the runtime distribution of all solutions for a given task. This method mitigates the impact of different hardware environments, allowing for fair cross-platform comparisons.
\end{itemize}

\subsection{Training Setup}

We sample and test 100 instances for each problem using Qwen-2.5-Coder-Instruct-7B to construct the SFT/DPO training data. For training data construction, we group multiple Abstract Syntax Tree (AST)-consistent code snippets based on their function ASTs, treating them as a set of homogeneous functions. From each such set, we randomly select one to serve as the response.

\begin{table}[h]
    \centering
    \begin{tabular}{lr}
        \toprule
        \textbf{Hyperparameter} & \textbf{Value} \\
        \midrule
        Learning rate & $5 \times 10^{-6}$\\
        Learning rate scheduler  & cosine \\
        Warmup ratio  & 0.03 \\
        Batch size    & 1024 \\
        Epoch         & 10   \\
        Data size     & 5000 \\
        \bottomrule
    \end{tabular}
    \caption{Hyperparameters For SFT}
    \label{tab:hyperparameters-sft}
\end{table}

\subsubsection{SFT} We evaluate accuracy and runtime efficiency using the sampled results. From each problem, we select the two correct samples with the highest runtime efficiency. Subsequently, 5,000 such samples are filtered to form the Supervised Fine-Tuning (SFT) training dataset. Training is conducted using LlamaFactory, \cite{zheng2024llamafactory}, and the hyperparameters used during training are presented in \Cref{tab:hyperparameters-sft}.

\begin{table}[h]
    \centering
    \begin{tabular}{lr}
        \toprule
        \textbf{Hyperparameter} & \textbf{Value} \\
        \midrule
        Learning rate & $5 \times 10^{-6}$\\
        Learning rate scheduler  & cosine \\
        Warmup ratio  & 0.03   \\
        Batch size    & 1024   \\
        Epoch         & 1      \\
        Data size     & 50000  \\
        \bottomrule
    \end{tabular}
    \caption{Hyperparameters For DPO}
    \label{tab:hyperparameters-dpo}
\end{table}

\subsubsection{DPO} We construct Correctness Pairs and Efficiency Pairs for DPO: 
(1) We begin by classifying the sampled program execution results. These are categorized into correct samples and incorrect samples, from which we then form the Correctness Pairs. 
(2) Building upon the correct samples, we next focus on their execution performance. We measure this based on the runtime CPU instruction count. 
Training is conducted using LlamaFactory, \cite{zheng2024llamafactory}, and the hyperparameters used during training are presented in \Cref{tab:hyperparameters-dpo}.

\begin{table}[h]
    \centering
    \begin{tabular}{lr}
        \toprule
        \textbf{Hyperparameter} & \textbf{Value} \\
        \midrule
        Episodes & 30 \\
        Learning rate & $1 \times 10^{-6}$\\
        Temperature & 1.0 \\
        Batch size & 64 \\
        Mini-batch size & 32 \\
        N (sample num) & 64 \\
        Max length & 16k \\
        Max generation length & 12k \\
        \bottomrule
    \end{tabular}
    \caption{Hyperparameters For RLOO/GRPO}
    \label{tab:hyperparameters-rl}
\end{table}

\subsubsection{RLOO/GRPO} For the RLOO/GRPO training process, we utilize all samples from the Train Dataset detailed in \Cref{tab:dataset}. Training is conducted using Verl~\cite{sheng2024hybridflow}, and the hyperparameters employed during this training process are presented in \Cref{tab:hyperparameters-rl}.

\subsection{GRPO with Round}

\begin{figure}[h]
    \centering
    \begin{subfigure}[b]{0.32\linewidth}
        \centering
        \includegraphics[width=\linewidth]{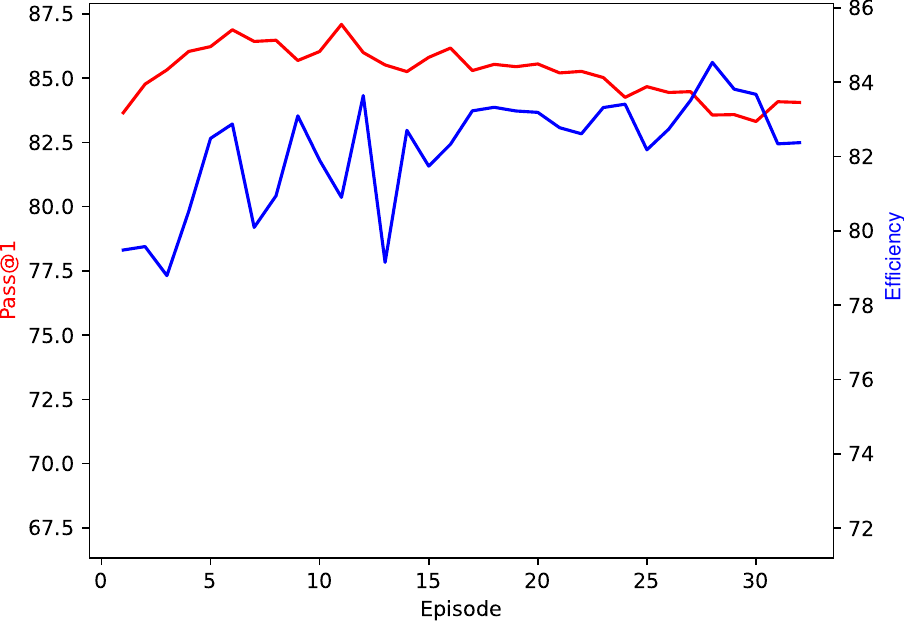}
        \caption{GRPO}
        \label{fig:grpo-dpo}
    \end{subfigure}
    \hfill
    \begin{subfigure}[b]{0.32\linewidth} 
        \centering
        \includegraphics[width=\linewidth]{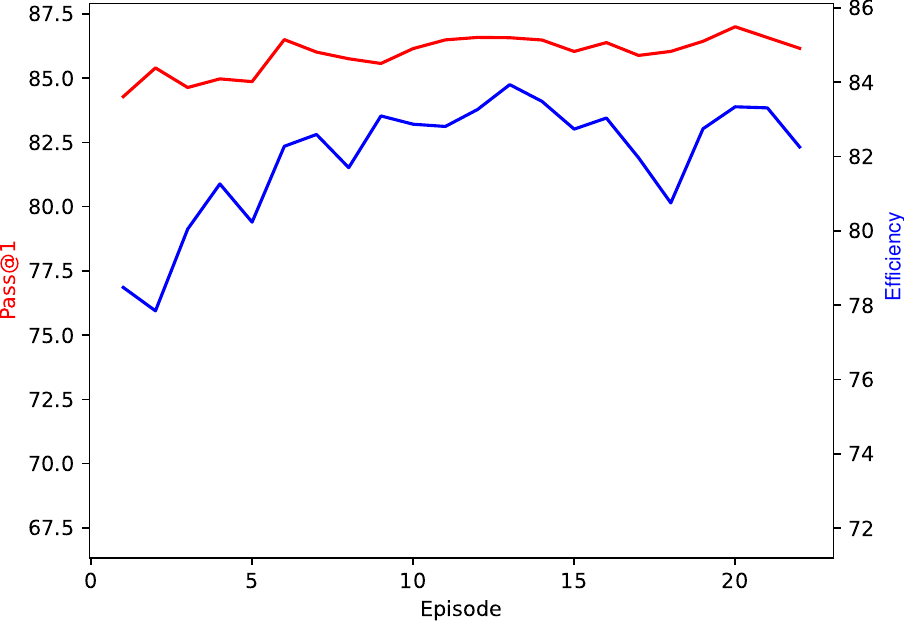}
        \caption{GRPO w/ Round}
        \label{fig:grpo-dpo-r1}
    \end{subfigure}
    \hfill
    \begin{subfigure}[b]{0.32\linewidth}
        \centering
        \includegraphics[width=\linewidth]{figures/episode/rloo-dpo.pdf}
        \caption{RLOO}
        \label{fig:rloo-dpo-2}
    \end{subfigure}
    \caption{
    Comparison of GRPO, GRPO with Round and RLOO performance in terms of accuracy and efficiency, initlized with the model after DPO. 
    The stability of GRPO's correctness is substantially improved by introducing a ``Round'' mechanism, making it similar to RLOO in its ability to maintain correctness. 
    This is a clear advantage over directly using the logarithmically scaled Instruction Count as the reward signal.
    }
    \label{fig:performance-var2}
\end{figure}

Compared to RLOO, GRPO's division by the standard deviation in the reward calculation can amplify small errors. This can be mitigated by introducing a rounding mechanism, with the experimental results presented in \Cref{fig:performance-var2}.

\end{document}